 \documentclass[11pt,reqno]{amsart}
 \usepackage{amsxtra}
 \usepackage[innercaption]{sidecap}
 \usepackage{graphicx}
 \graphicspath{{./images/}}
 \usepackage{amssymb}   
 \usepackage{enumitem}

\theoremstyle{plain}

\def\CC {{\mathbb C}}
\def\RR {{\mathbb R}}
\def\NN {{\mathbb N}}
\def\ZZ {{\mathbb Z}}
\def\PP {{\mathbb P}}

\def\be {\begin{eqnarray}}
\def\ben {\begin{eqnarray*}}
\def\ee {\end{eqnarray}}
\def\een {\end{eqnarray*}}

\def\AAA{\kern-0.3em}
\def\AA{\kern-0.18em}
\def\AC{\kern-0.14em}
\def\AB{\kern-0.22em}

\newcommand \nc {\newcommand}

\newtheorem{theorem}{Theorem}[section]
\newtheorem{lemma}[theorem]{Lemma}
\newtheorem{proposition}[theorem]{Proposition}
\newtheorem{corollary}[theorem]{Corollary}
\newtheorem{definition}[theorem]{Definition}
\newtheorem{example}[theorem]{Example}
\newtheorem{remark}[theorem]{Remark}
\nc \bth[1] { \begin{theorem}\label{t#1} } \nc \ble[1] {
\begin{lemma}\label{l#1} } \nc \bpr[1] {
\begin{proposition}\label{p#1} } \nc \bco[1] {
\begin{corollary}\label{c#1} } \nc \bde[1] {
\begin{definition}\label{d#1}\rm } \nc \bex[1] {
\begin{example}\label{e#1}\rm } \nc \bre[1] {
\begin{remark}\label{r#1}\rm } \nc \bcon[1] {
\medskip\noindent{\it{Conjecture #1}} } \nc \bqu[1]  {
\medskip\noindent{\it{Question #1}} }
\nc {\ethe} { \end{theorem} }
 \nc {\ele} { \end{lemma} } \nc {\epr}
{ \end{proposition} } \nc {\eco} { \end{corollary} } \nc {\ede} {
\end{definition} } \nc {\eex} { \end{example} } \nc {\ere} {
\end{remark} } \nc {\econ} {\smallskip} \nc {\equ} {\smallskip}
 \nc \thref[1]{Theorem \ref{t#1}}
\nc \leref[1]{Lemma \ref{l#1}} \nc \prref[1]{Proposition
\ref{p#1}} \nc \coref[1]{Corollary \ref{c#1}} \nc
\deref[1]{Definition \ref{d#1}} \nc \exref[1]{Example \ref{e#1}}
\nc \reref[1]{Remark \ref{r#1}}

\def \D {{\mathcal D}}

\def \B {{\mathcal B}}
\def \T {{\mathcal T}}
\def \L {{\mathcal L}}

\def \diag { {\mathrm{diag}} }

 \def\AA  {\kern-0.1em}
 \def\BB  {\kern+0.1em}
 \def\BBB {\kern+0.15em}
 \def\K   {\kern+0.05em}
 \def\MK  {\kern-0.07em}
 \def\MKK {\kern-0.04em}

\begin{document}

\vspace{0.5cm}

\title[ Non-integrability of the Painlev\'{e} IV ]
{ Non-integrability of the Painlev\'{e} IV equation in the Liouville-Arnold sense and Stokes phenomena  }

\author[Tsvetana  Stoyanova]{Tsvetana  Stoyanova}

\date{27.02.2023}

 \maketitle

\begin{center}
{Department of Mathematics and Informatics,
Sofia University,\\ 5 J. Bourchier Blvd., Sofia 1164, Bulgaria, 
cveti@fmi.uni-sofia.bg}
\end{center}

\vspace{0.5cm}

\begin{abstract}
	In this paper we study the integrability of the Hamiltonian system associated to the fourth Painlev\'{e} equation.
	We prove that one two parametric family of this Hamiltonian system is non integrable in the sense of the
	Liouville-Arnold theorem. Computing explicitly the Stokes matrices and the formal invariants of the second variational equations
	 we  deduce that the connected component of the unit element of 
	corresponding differential Galois group is not Abelian. Thus the Morales-Ramis-Sim\'{o} theory leads to a non-integrable result.
	Moreover, combining the new result with our previous one we establish that for all values of the parameters for which
	the $P_{IV}$ equation has a particular rational solution the corresponding Hamiltonian system is not integrable
	by meromorphic first integrals which are rational in $t$.
	 \end{abstract}



\headsep 10mm \oddsidemargin 0in \evensidemargin 0in

\section{Introduction}

 In this paper we study the integrability of the fourth Painlv\'{e}  equation $P_{IV}$
  \be\label{P4}
   \ddot{y}=\frac{1}{2 y}\,(\dot{y})^2 + \frac{3}{2}\,y^3 + 4\,t\,y^2 + 2 (t^2-a)\,y + \frac{b}{y},\quad
   \cdot=\frac{d}{d t}\,,
  \ee
  where $t\in\CC$ and $a, b$ are arbitrary complex parameters from the point of view of the Hamiltonian dynamics. 
  We prove that for one family of the parameters $a$ and $b$  the Painlev\'{e} IV equation is not integrable by
  meromorphic functions which are rational in $t$.
  
 There are many works devoted to both pure mathematical aspects of the $P_{IV}$ equation  
   \cite{BM, BRG, CF, C1, DK, FvAZ, JR, NO, NY, O} and to its applications to  mathematical physics   
 \cite{FIK, BCH1, C}.  
 In the present paper we are concerned with the Hamiltonian system corresponding to the $P_{IV}$ equation.
 We prove rigorously that for one two parametric family of $P_{IV}$ equation this Hamiltonian system  is not integrable,
 i.e. there no exist two independent meromorphic first integrals in involution.
  As in our previous works \cite{HS, St1, St2, St3} our approach  is based 
  on the Morales-Ramis-Sim\'{o} theory
  which reduces the problem of integrability of a given Hamiltonian system to the problem of 
  determination of the differential Galois group of the variational equations along a particular non-equilibrium solution.
  In the main part of this paper we deal with the integrability of the equation \eqref{P4} when $a=0, b=-\frac{2}{9}$ and
  $y=-\frac{2}{3}\,t$. It urns out that the differential Galois group of the first variational equations is Abelian.
  Then according to the scheme of Morales-Ramis-Sim\'{o} we have to consider the higher variational equations.
  The second normal variational equations $(\textrm{NVE})_2$ are reduced to a fourth-order linear reducible ordinary differential
  equations with two singular points taken at the origin and the infinity point. The origin is a non-resonant
  irregular point of Poincar\'{e} rank 1 while the infinity point is a resonant Fuchsian singularity. 
  By an explicit computation we show that the local fundamental set of solutions at the origin of the
  $(\textrm{NVR})_2$ contains the functions
   $$\,
        \phi_{\theta}(z)=3\,(-i\,\sqrt{3})^{-5/2}\int_0^{+\infty\,e^{i \theta}}
        \frac{e^{-\frac{\xi}{z}}}{\left(\xi -i\,\sqrt{3}\right)^{5/2}}\,d \xi 
    \,$$
    and
    $$\,    
          \varphi_{\theta}(z)=3\,(i\,\sqrt{3})^{-5/2}\int_0^{+\infty\,e^{i \theta}}
          \frac{e^{-\frac{\xi}{z}}}{\left(\xi + i\,\sqrt{3}\right)^{5/2}}\,d \xi\,.
   \,$$
  The analytic continuation of these functions leads to the existence of non-trivial Stokes matrices at the origin of the
  $(\textrm{NVE})_2$ and therefore  implies
  non-commutative connected component of the unit element of the corresponding differential Galois group.
  Then the key theorem  (\thref{b} in the Section 3) of this paper states

    \bth{key}
    Assume that  $a=0$ and $b=-\frac{2}{9}$. Then the fourth Painlev\'{e} equation is not integrable by meromorphic
    functions which are rational in $t$.
    \ethe
    
    From the paper of Murata \cite{M} it follows that all the rational solutions of the $P_{IV}$
    equation \eqref{P4} can be obtained by B\"{a}cklund transformations from the following two particular
    solutions
    \ben
    (\textrm{I})\quad & & v_1=v_2=v_3=0 \quad (\textrm{respectively} \quad a=1, b=0 ), \quad
    y=p=0,\\[0.1ex]
    (\textrm{II})\quad & & v_1=-v_3=\frac{1}{3}, v_2=0 \quad (\textrm{respectively} \quad a=0, b=-\frac{2}{9} ), \quad
    y=-p=-\frac{2}{3}\,t\,.
    \een
    Since the  B\"{a}cklund transformations are birational canonical transformations \cite{O, Wa, ZF} 
    we can extend the result of \thref{key} to the main  theorem of this paper

    \bth{main}
    Assume that 
    $$\,
    a=m,\quad b=-\frac{2}{9}\,(1+3 n)^2\,,
    \,$$
    where $m, n\in\ZZ$ such that $\frac{1}{2} (m+n)\in\ZZ$. Then the fourth Painlev\'{e}  equation \eqref{P4} is   
    not integrable in the Liouville-Arnold sense by meromorphc functions that are rational in $t$.
    \ethe
    
    In \cite{St3} we have proved that when the $P_{IV}$ equation has a rational solution which is obtained by
    B\"{a}cklund transformations from the solution (I) the Hamiltonian system \eqref{H} is not integrable in the Liouville-Arnold
    sense. 
    Combining \thref{main} and Theorem 1.1 in \cite{St3} we establish
    
    \bth{ex}
    The fourth Painlev\'{e} equation \eqref{P4} is not integrable in the Louville-Arnold sense by meromorphic
    functions which are rational in $t$.
    \ethe

   Noumi and Okamoto are the first authors who pose the problem of irreducibility of the fourth Painlev\'{e} equation from
   a strictly mathematical point of view. In \cite{NO} they prove that except some special particular solutions 
   (rational, algebraic, Riccati solutions), any solution of the $P_{IV}$ equation is non-classical in the sense of Umemura.
   In \cite{ZF} Zoladek and Filipuk prove non-integrability of the fourth Painlev\'{e} equation as a Hamiltonian system by
   algebraic first integrals. Their approach is based on the Liouville's theory of elementary functions and some
   properties of elliptic integrals. In \cite{AHPT} Acosta-Hum\'{a}nez, van der Put and Top claim that when $a=0, b=-\frac{2}{9}$
   and $a=0, b=-2$ the Painlev\'{e} IV equation is not integrable as a Hamiltonian system. Their approach is based
   on the Morales-Ramis-Sim\'{o} theory. In particular, they like us study the
   differential Galois group of the second normal variational equations. Then their statement follows from the fact 
   that the solution of the second variational equations contains
   non-trivial functions, like error function, which implies non-commutative Galois group and therefore non-integrability.
   We note that the authors do not build explicitly the solution of the second variational equations, as well as
   they do not compute generators of the corresponding differential Galois group. Our method show  in details   
   the analytic obstruction for integrability and present explicitly the generators of the differential Galois group
   of the $(\textrm{NVE})_2$.

       This paper is organized as follows. In the next section we briefly review the basics of the Morales-Ramis-Sim\'{o}
       theory of the non-integrability of the Hamiltonian systems, as well as of the theory of linear ordinary
       differential equations and its relation to the differential Galois theory. We also recall the Hamiltonian system 
       corresponding to the $P_{IV}$ equation. 
      In Section 3 we study for integrability the Hamiltonian system \eqref{H} when $a=0$
       and $b=-2/9$.
       The main result of this section is \thref{b} which states that when $a=0, b=-2/9$ the fourth Painlev\'{e} equation \eqref{P4}
       is not integrable in the Louville-Arnold sense by meromorphic functions which are rational in $t$.
       In Section 4 we prove \thref{main}.
    
%
    
    \section{Preliminaries}

    \subsection{Non-intagrability and differential Galos theory}

    In this paragraph we briefly recall the theory of Morales-Ruiz, Ramis and Sim\'{o} about non-integrability of
    Hamiltonian systems following \cite{MR, MRS}.
    
    Let $M$ be a symplectic analytical complex manifold of complex dimension $2 n$. Consider on $M$ a Hamiltonian system
     \be\label{Ham}
      \dot{x}=X_H(x)
     \ee
    with a Hamiltonian $H : M \rightarrow \CC$. Recall that by the theorem of Liouville-Arnold \cite{Ar} the Hamiltonian system \eqref{Ham}
    is completely integrable in the Louville-Arnold sense if there exist $n$ first integrals
    $f_1=H, f_2, \ldots, f_n$ functionally independent and in involution. 
   Let $x(t)$ be a particular solution of \eqref{Ham} which is not an equilibrium point of the Hamiltonian vector field $X_H$. Denote by
   $\Gamma$ the phase curve corresponding to this solution. Then the first variational equations along $\Gamma$ are given by
    \be\label{ve}
     \dot{\xi} = \frac{\partial X_H}{\partial x} (x(t))\,\xi\,.
    \ee    
    Using the Hamiltonian $H$ we can reduce the variational equation \eqref{ve} in the following sense. Consider the normal bundle
    of $\Gamma$ on the level variety $M_h=\{x\,|\,H(x)=h\}$. The projection of the variational equation \eqref{ve} on this
    bundle is the so called normal variational equation $(\textrm{NVE})$. The dimension of the $\textrm{NVE}$ is $2 n-2$.
    The solutions of the $\textrm{NVE}$ define a Picard-Vessiot extension $L_1$ of the differential field $K$ of the coefficients of
    $\textrm{NVE}$. This in its turn defines a differential Galois group $G_1=Gal(L_1/K)$.  
    Then the main theorem of the Morales-Ruiz and Ramis theory states \cite{MR, MRS}
    
    \bth{non-1}{\bf $($Morales-Ruiz and Ramis$)$}
      If the Hamilton system \eqref{Ham} is completely integrable with meromorphic first integrals in a
      neighbourhood of $\Gamma$, not necessarily independent on $\Gamma$ itself, then the identity component $G_1^0$ of
      the differential Galois group $G_1=Gal(L_1/K)$  is Abelian.
    \ethe
    
   The opposite is not true in general, i.e. if the connected component of the unit element of the differential
   Galois group $Gal (L_1/K)$ is Abelin, it is not sure that the corresponding Hamiltonian system \eqref{Ham} is integrable.
    To overcome this problem Morales-Ruiz, Ramis and Sim\'{o} propose to use higher order variational equations \cite{MRS}.
    Let again $x(t)$ be a particular solution of \eqref{Ham} which is not an equilibrium point of the vector field $X_H$.
    We write the general solution as $x(t, z)$, where $z$ parametrizes it near $x(t)$ as $x(t, z_0)=x(t)$. Then we write 
    the system \eqref{Ham} as
     \be\label{Ham-1}
      \dot{x}(t, z)=X_H(x(t, z))\,.
     \ee
    Denote by $x^{(1)}(t, z), x^{(2)}(t, z), \ldots $ the derivatives of $x(t, z)$ with resect to $z$ and by\\
    $X_H^{(1)}(x), X_H^{(2)}(x), \ldots$ the derivatives of $X_H(x)$ with respect to $x$. By successive derivation of
    \eqref{Ham-1} with respect to $z$ and evaluation at $z_0$ we obtain the so called $k$-th variational equations
    $(\textrm{VE})_k$ for the function $x(t)$
     \be\label{vek}\qquad
      \dot{x}^{(k)}(t)=X_H^{(1)}(x(t))\,x^{(k)}(t) + P\left(x^{(1)}(t), x^{(2)}(t), \ldots, x^{(k-1)}(t)\right)\,.
     \ee    
   Here $P$ is a polynomial. The coefficients depend on $t$ through $X_H^{(j)}(x(t), j < k$. For every $k > 1$ the linear non-homogeneous
   system \eqref{vek} can be arranged as a linear homogeneous system of higher dimension. This chain of
  linear homogeneous systems defines a chain of  successive Picard-Vessiot extensions of the main differential field $K$ of the
  coefficients of $\textrm{NVE}$, i.e. we have
  $K \subset L_1 \subset L_2 \subset \cdots \subset L_k$, where $L_1$ is as above, $L_2$ is the Picard-Vessiot extension of $K$
  associated with $(\textrm{VE})_2$, etc. 
  We can define the Galois groups $G_1=Gal(L_1/L), G_2=Gal(L_2/K), \ldots$. Then the main theorem in the Morales-Ramis-Sim\'{o}
  theory states \cite{MRS}
  
   \bth{non-h}{\bf $($Morales-Ruiz, Ramis and Sim\'{o}$)$}
    If the Hamiltonian system \eqref{Ham} is completely integrable then for every $m\in\NN$ the connected component of
    the unit element $G_m^0$ of the Galois group $G_m=Gal(L_m/K)$ is Abelian.
   \ethe
   Assume that $G_1^0$ is Abelian.
   Then the theorem \thref{non-h} says that if we want to obtain a non-integrable result we have to find a group
   $G_m^0, m > 1$ which is not Abelian. Note that this group will be a solvable group (see \cite{MRS} for details).
   In this case the non-integrability in the sense of Hamiltonian dynamics corresponds to integrability in the
   Picard-Vessiot sense.

     \subsection{The fundamental matrix solution of the $(\textrm{NVE})_2$}
     
      The main result of this paper is based on the study of the differential Galois group of the second normal variational
      equations since the differential Galois group of the first variational equations is Abelian.
     Throughout this paper we assume that $K=\CC(z)$. The Cyclic Vector Theorem ensures that
    the homogeneous  linear system associated with $(\textrm{NVE})_2$ can be always reduced to a  
    higher order linear scalar equation. As the initial linear system and the obtained 
    scalar equation belong to the same differential module the connected components $G^0_2$ of the corresponding differential Galois groups
    are Abelian or not-Abelian at the same time (see \cite{Si, vPS} for more details).
    It turns out that the $(\textrm{NVE})_2$ are reduced to a fourth order reducible scalar equation.
    For this reason from here to the end of this section we 
       consider a reducible fourth order linear ordinary differential equation in the form
       \be\label{r}
       L\,y=\left(L_2 \circ L_1\right)\,y=0,\quad L_2 \circ L_1 \neq L_1 \circ L_2\,,
       \ee  
       where $L_j, j=1, 2$ are second order differential operators
       \be\label{op}
       L_1=\partial^2 + a_1(z)\,\partial + a_0(z),\quad
       L_2=\partial^2 + b_1(z)\,\partial + b_0(z)
       \ee       
       with $\partial=\frac{d}{d z}$ and coefficients $a_j(z), b_j(z)\in\CC(z)$. 
       For simplicity throughout this section we call the equation \eqref{r} with the operators \eqref{op}
       just the $(\textrm{NVE})_2$.
    
    Denote by $\{w_1(z), w_2(z)\}$ a fundamental set of solutions of the equation
 $L_2\,y=0$. Then
 
   \bth{fss}
   With the above notatons
     the equation \eqref{r}  possesses a fundamental set of solutions
     $\{y_1(z), y_2(z), \psi_1(z), \psi_2(z)\}$ where $\{y_1(z), y_2(z)\}$ is a fundamental set of solutions of
     the equation $L_1\,y=0$. The functions $\psi_1(z)$ and $\psi_2(z)$ are the solutions of the equations
     $L_2\,y=w_j(z), j=1, 2$, respectively.
   \ethe
   
   \proof
   
     The proof is straightforward.
     \qed
     
   Denote by $w(z)$ the solution of the equation $L_2\,y=0$.
   The next theorem associate with the equation \eqref{r}  a fourth order linear system
   of differential equations. 
 
  \bth{e-s}
  The function $y(z)$ solves the equation \eqref{r}  if and only if the vector 
  $Y(z)=(y(z), y'(z), w(z), w'(z))^{\tau}$ solves the linear system
   $$\,
    Y'(z)=A(z)\,Y(z), \quad '=\frac{d}{d z}\,,
   \,$$
  where
   \ben
     A(z)=\left(\begin{array}{cc}
     	A_1(z)   & B\\
     	0        & A_2(z)
     	   \end{array}\right)\,. 
    \een
    The matrices $A_j(z),j=1, 2$ and $B$ are second order matrices in the form
     \ben
      A_1(z)=\left(\begin{array}{cc}
      	 0        & 1\\
      	-a_0(z)   & -a_1(z)
      \end{array}\right),\quad
       A_2(z)=\left(\begin{array}{cc}
       	0         & 1\\
       	-b_0(z)   & -b_1(z)
       \end{array}\right),\quad
       B=\left(\begin{array}{cc}
       	0   & 0\\
       	1   & 0
       	\end{array}\right)\,.
       	     \een
  \ethe
  
  \proof
  
  The proof is straightforward.
  \qed
  
  \bth{fms}
   The equation \eqref{r} possesses a fundamental matrix solution $\Phi(z)$ in the form
    \be\label{fss}
     \Phi(z) =\left(\begin{array}{cc}
     	 \Phi_1(z)  & \Psi(z)\\
     	   0        & \Phi_2(z)
     	\end{array}\right)\,,
    \ee
    where $\Phi_j(z), j=1, 2$ are fundamental matrix solutions of the second order linear systems
    $U'(z)=A_j(z)\,U(z)$, respectively, with matrices $A_j(z), j=1, 2$ defined by \thref{e-s}.
    The matrix $\Psi(z)$ writes
     \be\label{n}
      \Psi(z)=\left(\begin{array}{cc}
      	 \psi_1(z)    & \psi_2(z)\\
      	 \psi'_1(z)   & \psi'_2(z)
      	\end{array}\right)\,,
     \ee
     where $\psi_j(z) j=1, 2$ are introdused by \thref{fss}.
    \ethe
  
  \proof
  
    The proof is straightforward.
    \qed
  
   \bre{m}
    The matrices $\Phi_j(z), j=1, 2$ from \thref{fss} appear as fundamental matrix solutions for the equations 
    $L_j\,y=0, j=1, 2$, respectively.
   \ere

     \subsection{The  differential Galois group of the $(\textrm{NVE})_2$}

   In this paragraph we review some definitions, facts and notation from the theory of linear ordinary differential equations,
   as well as, from the Picard-Vessiot theory which are required to describe the differential Galois group of 
   one equation of the kind \eqref{r}. Most of them can be found in \cite{R2, Si, vPS}.         
     Throughout this paragraph we assume that the equation \eqref{r} 
     has only two singular points over $\CC\PP^1$: one non-resonant irregular singularity
     at the origin of Poincar\'{e} rank 1 and one regular singularity at $z=\infty$. 
     
     We start by describing the local differential Galois group at the origin.
      The restriction of the theorem of Hukuhara-Turrittin-Wasow \cite{W} gives us
     
     \bth{formal-0}
      Assume that the equation \eqref{r}  has a non-resonant singularity at the origin.
      Then the fundamental matrix solution $\Phi(z)$ from \eqref{fss} is represented at the origin as
       \be\label{fo}
        \hat{\Phi}_0(z)=\hat{H}(z)\,z^{\Lambda}\,\exp \left(\frac{Q}{z}\right)\,,
       \ee
       where $\Lambda$ and $Q$ are constant diagonal matrices. The entries of the matrix $\hat{H}(z)$ are usually 
       divergent power series.
     \ethe
     The formula \eqref{fo} is usually regarded as a formal fundamental matrix solution at the origin.
     To introduce the formal invariants of the equation \eqref{r} we consider the equation \eqref{r} and its
     formal fundamental matrix solution $\hat{\Phi}(z)$ from \eqref{fo} over the field $\CC((z))$ of formal power series
     in $z$.
     \bde{mon}
     {\it  With respect to the formal fundamental matrix solution $\hat{\Phi}_0(z)$ given by \eqref{fo} we define the formal monodromy
      matrix $\hat{M}_0\in GL_4(\CC)$ around the origin as
       $$\,
        \hat{\Phi}_0(z.e^{2 \pi\,i})=\hat{\Phi}_0(z)\,\hat{M}_0\,.
       \,$$
       In particular,
       $$\,
        \hat{M}_0=e^{2 \pi\,i\,\Lambda}\,.
       \,$$}
     \ede
      Denote $\Lambda=\diag(\lambda_1, \lambda_2, \lambda_3, \lambda_4)$ and $Q=\diag (q_1, q_2, q_3, q_4)$.
      \bde{ext}
       {\it With respect to the formal fundamental matrix solution $\hat{\Phi}_0(z)$ given by \eqref{fo} we define the exponential torus
       $\T$ as the differential Galois group $Gal(E/F)$, where
       $F=\CC((z))(z^{\lambda_1}, z^{\lambda_2}, z^{\lambda_3}, z^{\lambda_4})$ and $E=F(e^{q_1/z}, e^{q_2/z}, e^{q_3/z}, e^{q_4/z})$.} 
      \ede       
     Since for $\sigma\in\T$ we have that $\sigma(z^{\lambda})=z^{\lambda}$ and $\sigma(e^{q/z})=c\,e^{q/z},\,c\in\CC^*$ we can
     consider $\T$ as a subgroup of $(\CC^*)^4$. The formal monodromy $\hat{M}_0$ and the exponential torus $\T$ generate topologically
     the differential Galois group at the origin of the equation \eqref{r} over $\CC((z))$ (see Theorem 1.4.9 in \cite{Si}).

       To introduce the analytic invariants at the origin we consider the equation \eqref{r} and its solutions over the field
       $\CC(\{z\})$ of convergent power series in $z$.
      A fourth order linear ordinary differential equation  \eqref{r} with a non-resonant irregular singularity at the 
     origin of Poincar\'{e} rank 1 admits a local fundamental set of solutions at the origin in the form
      \ben
         & &
       y_1(z)=z^{l_1}\,e^{\frac{q_1}{z}}\,\hat{\varphi}_1(z),\quad
       y_2(z)=z^{l_2}\,e^{\frac{q_2}{z}}\,\hat{\varphi}_2(z),\\[0.1ex]
         & &
         y_3(z)=z^{l_3}\,e^{\frac{q_3}{z}}\,\hat{\varphi}_3(z),\quad
         y_4(z)=z^{l_4}\,e^{\frac{q_4}{z}}\,\hat{\varphi}_4(z)\,,
      \een
      where $q_j\in\CC, 1 \leq j\leq 4$ are as above. The numbers $l_j$ and $\lambda_j$ from above are related by $\lambda_j-l_j\in\ZZ$. Here $\hat{\varphi}_j(z), 1 \leq j \leq 4$ are power series in $z$
      which are either convergent or divergent. 
      
      \bde{sd}
       {\it Under the above notations for every divergent power series $\hat{\varphi}_j(z)$ we define a family $\Theta_j$ of
       admissible singular directions 
        $$\,
         \Theta_j=\{\theta_{ji},\,0 \leq \theta_{j i} < 2 \pi\}
        \,$$
        where $\theta_{j i}$ is the bisector of the sector  $\{Re \left(\frac{q_i-q_j}{z}\right) < 0\}$. In particular,
        $$\,
         \Theta_j=\left\{\theta_{j i},\, 0 \leq \theta_{j i} < 2 \pi,\,
         \theta_{j i}=\arg (q_j-q_i),\, 1 \leq i \leq 4, i\neq j \right\}\,.
        \,$$}
          \ede
    
    The application of the summability theory to the linear ordinary equations leads to the following important theorem
    of Ramis \cite{R2}
    
    \bth{R}
     In the formal fundamental matrix solution $\hat{\Phi}_0(z)$ from \eqref{fo} the entries of the matrix $\hat{H}(z)$
     are 1-summable in every non-singular direction $\theta$.
     If we denote by $H_{\theta}(z)$ the 1-sum of the matrix $\hat{H}(z)$  along a non-singular direction
     $\theta$ then the matrix $\Phi_0^{\theta}(z)=H_\theta(z)\,z^{\Lambda}\,\exp\left(\frac{Q}{z}\right)$
     gives an actual fundamental matrix solution of the equation \eqref{r} on a small sector bisected by $\theta$.
    \ethe

    For the needed aspects of the summability theory we refer to the works of Loday-Richaud \cite{LR}, as well as the works of 
    Ramis \cite{R1, R2}.

    Let $\epsilon > 0$ be a small number. Let $\theta-\epsilon$ and $\theta+\epsilon$ be tow non-singular neighboring
    directions of the singular direction $\theta\in\Theta_j$. Let $\Phi_0^{\theta+\epsilon}(z)$ and $\Phi_0^{\theta-\epsilon}(z)$
    be the actual fundamental matrix solutions at the origin of the equations \eqref{r} related to the direction
    $\theta+\epsilon$ and $\theta-\epsilon$ in the sense of \thref{R}. Then
    
    \bde{stokes}
     {\it With respect to the actual fundamental matrix solutions $\Phi_0^{\theta-\epsilon}(z)$ and $\Phi_0^{\theta+\epsilon}(z)$
     the Stokes matrix $\textrm{St}_{\theta}\in GL_4(\CC)$ related to the singular direction $\theta$ is defined as
      $$\,
       \textrm{St}_{\theta}=\left(\Phi_0^{\theta+\epsilon}(z)\right)^{-1}\,\Phi_0^{\theta-\epsilon}(z)\,.
      \,$$}
    \ede  
    The next theorem of Ramis \cite{R3} describes the differential Galois group at the origin of the
    equation \eqref{r}.
    
    \bth{Ramis}{\bf$($Ramis$)$}
    The differential Galois group at the origin of the equation \eqref{r} is the Zariski closure in
    $GL_4(\CC)$ of the group generated by the formal monodromy $\hat{M}$, the exponential torus $\T$ and the Stokes
    matrices $St_\theta$ for all singular directions $\theta$.
    \ethe
  
  The differential Galois group
    at $z=\infty$ of the equation \eqref{r} is computed by the following theorem of Schlesinger \cite{vPS}
    
   \bth{Sh}{\bf $($Schlesinger$)$}
    The differential Galois group at a regular singularity is the Zariski closure of the monodromy group around
    this point.
    \ethe
     For the needed facts of the local theory around regular singular points we refer to the book of Golubev \cite{G}.
     
     In Proposition 1.3 in \cite{Mi} Mitschi describes the global differential Galois group of
     an arbitrary differential equation. Here we formulate the result of Mitschi for the equation \eqref{r}.
     
     \bth{Mitschi}{\bf $($Mitschi$)$}
      The global differential Galois group $G$ of the equation \eqref{r} is topologically generated in $GL_4(\CC)$ by the 
      differential Galois group at the origin and the differential Galois group at $z=\infty$ of the same equation. 
     \ethe

   \subsection{$P_{IV}$ equation as a Hamiltonian system}
     
       The fourth Painlev\'{e} equation \eqref{P4} is equivalent to the following non-autonomous
       Hamiltonian system of $1+1/2$ degrees of freedom \cite{NY, O}
       $$\,
       \dot{y}=\frac{\partial H_{IV}}{\partial p},\quad
       \dot{p}=-\frac{\partial H_{IV}}{\partial y}
       \,$$
       with the polynomial Hamiltonian \cite{NY}
       $$\,
       H_{IV}=y\,p^2 -y^2\,p -2 t\,y\,p - 2 (v_1-v_2)\,p - 2 (v_2-v_3)\,y\,.
       \,$$
       The parameters $v_1, v_2, v_3$ satisfy the condition $v_1+v_2+v_3=0$. They and the parameters $a, b$ in
       \eqref{P4} are related through the formulas \cite{NY}
       $$\,
       a=1 + 3 v_3,\qquad b=-2 (v_1-v_2)^2\,.
       \,$$
       We can extend this Hamiltonian system to an autonomous Hamiltonian system on $\CC^4$ by introducing two new dynamical
       variables $t$ and $F$. The variable $F$ is conjugate to $t$. The new obtained Hamiltonian system is
       already autonomous of  two degrees of freedom with Hamiltonian
       $H=H_{IV} +F$. The new Hamiltonian system becomes
       \be\label{H}
       & &
       \frac{d y}{d s}=\frac{\partial H}{\partial p}, \qquad
       \frac{d t}{d s}=\frac{\partial H}{\partial F},\\[0.15ex]
       & &
       \frac{d p}{ d s}=-\frac{\partial H}{\partial y},\qquad
       \frac{d F}{d s}=-\frac{\partial H}{\partial t}\,.\nonumber 
       \ee
       The symplectic  form $\Omega$ is canonical in the variables $y, p, t $ and $F$, i.e.
       $\Omega=d p \wedge d y + d F \wedge d t$.         
     
%
    
  \section{Non-integrability of the Painlev\'{e} IV for $a=0,\,b=-\frac{2}{9}$}
  
  In this section we prove that when $a=0$ and $b=-\frac{2}{9}$ the Hamiltonian system \eqref{H} is not-integrable in
  the Liouville-Arnold sense.
  
  \subsection{Hamiltonian system and the first variational equations}

  When $v_1=-v_3=\frac{1}{3},\,v_2=0$ the Hamiltonian system \eqref{H} becomes
   \be\label{h}
    \dot{y} &=& 2\,y\,p - y^2 -2\,t\,y - \frac{2}{3}\nonumber\\[0.15ex]
    \dot{p} &=& -p^2 + 2\,y\,p + 2\,y\,p + \frac{2}{3}\\[0.15ex]
    \dot{t} &=& 1\nonumber\\
    \dot{F} &=& 2\,y\,p\,.\nonumber 
   \ee  
The non-equilibrium particular solution along which we write the variational equations is
 \be\label{ps}
 y=-p=-\frac{2}{3}\,t,\quad t=s, \quad F=-\frac{8}{27}\,t^3.
 \ee
 Since $\dot{t}=1$ from here on we use $t$ as an independent variable instead of $s$.
 
  \bpr{p1}
   The differential Galois group of the first normal variational equations along the solution 
   \eqref{ps} is an Abelian group.
    \epr
  
   \proof
   The first normal variational equations $(\textrm{NVE})_1$ along the solution \eqref{ps} writes
    $$\,
     \dot{y}_1=\frac{2}{3}\,t\,y_1 - \frac{4}{3}\,t\,p_1,\quad
     \dot{p}_1=\frac{4}{3}\,t\,y_1 - \frac{2}{3}\,t\,p_1\,.
    \,$$
   The $NVE_1$ are solvable by quadratures and the matrix
    $$\,
     \Phi(t)=\left(\begin{array}{cc}
      2    & 2\\[0.15ex]
      1-i\,\sqrt{3}   & 1+i\,\sqrt{3}
             \end{array}\right)\,
    \left(\begin{array}{cc}
      e^{\frac{i\,\sqrt{3}\,t^2}{3}}    & 0\\[0.15ex] 
       0                                & e^{-\frac{i\,\sqrt{3}\,t^2}{3}}
       \end{array}\right)\,        
    \,$$
    is a fundamental matrix solution. Thus the differential Galois group of $\textrm{NVE}_1$ is isomorphic
    to the multiplicative group whch is an Abelian group.
    
    This ends the proof. 
       \qed
       
     \subsection{The second variational equations}  
       
       For the second normal variational equations $(\textrm{NVE})_2$ we obtain the fifth-order linear system
       \ben
        \dot{y}_2  &= & \frac{2}{3}\,t\,y_2 - \frac{4}{3}\,t\,p_2 + 2\,u - v\\[0.15ex]
        \dot{p}_2  &=& \frac{4}{3}\,t\,y_2 - \frac{2}{3}\,t\,p_2 + 2\,u - w\\[0.15ex]
        \dot{u}    &=& \frac{4}{3}\,t\,v - \frac{4}{3}\,t\,w\\[0.15ex]
        \dot{v}    &=& -\frac{8}{3}\,t\,u + \frac{4}{3}\,t\,v\\[0.15ex]
        \dot{w}    &=& \frac{8}{3}\,t\,u - \frac{4}{3}\,t\,w\,,
       \een
       where we have put $u=y_1\,p_1,\,v=y_1^2,\,w=p_1^2$. The $(\textrm{NVE})_2$ are reduced to the following
       fourth-order equation
        \be\label{nve2}
         \ddddot{y}-\frac{4}{t}\,\dddot{y}_2 + \frac{20}{3}\,t^2\,\ddot{y}_2
         -4\,t\,\dot{y}_2 - \left(12 - \frac{64}{9}\,t^4\right)\,y=0\,.
        \ee
        The transformation $t^2=1/z$ takes the equation \eqref{nve2} into the equation 
         \be\label{eq1}\qquad\qquad
          \frac{d^4 y}{d z^4} + \frac{11}{z}\,\frac{d^3 y}{d z^3} +
          \left(\frac{111}{4 z^2} + \frac{5}{3 z^4}\right)\,\frac{d^2 y}{d z^2}
          +\left(\frac{27}{2 z^3} + \frac{3}{z^5}\right)\,\frac{d y}{d z} 
          - \left(\frac{3}{4 z^6} - \frac{4}{9 z^8}\right)\,y=0\,.
         \ee
          The transformation $t^2=1/z$ changes the differential Galois group $G$ of the equation \eqref{nve2} but it
          preserves the connected component of the unit element of $G$.
         The equation \eqref{eq1} has two singular points over $\CC\PP^1$ --  the points $z=0$ and $z=\infty$.
         The origin is an irregular singularity of Poincar\'{e} rank 1, while $z=0$ is a regular singularity.

         Let us firstly determine the local differential Galois group at the origin of equation \eqref{eq1}. 
         The equation \eqref{eq1} is a reducible equation and it can be present as 
         $L\,y=(L_2\circ L_1) \,y=0,\,$ where the operators $L_2$ and $L_1$ do not commute and
          $$\,
           L_2=\partial^2 + \frac{9}{z}\,\partial + \frac{55}{4 z^2} + \frac{4}{3 z^4},\qquad
           L_1=\partial^2 + \frac{2}{z}\,\partial + \frac{1}{3 z^4},\quad
           \partial=\frac{d}{d z}\,.
          \,$$
          Every equation $L_j\,y=0, j=1, 2$ is solvable. The system of functions 
          \be\label{sol}
           y_1(z)=\exp \left(\frac{i\,\sqrt{3}}{3 z}\right),\quad 
           y_2(z)=\exp \left(\frac{-i\,\sqrt{3}}{3 z}\right) 
          \ee
          is a fundamental set of solutions of the equation $L_1\,y=0$ while the system 
            \be\label{sol-1}
           w_1(z) &=&
           z^{-11/2}\,\exp \left(\frac{2 i\,\sqrt{3}}{3 z}\right)\,\left[z^2 + \frac{i\,\sqrt{3}}{2}\,z^3\right],\\[0.2ex] 
           w_2(z) &=& 
           z^{-11/2}\,\exp \left(\frac{-2 i\,\sqrt{3}}{3 z}\right)\,\left[z^2 - \frac{i\,\sqrt{3}}{2}\,z^3\right] 
           \nonumber 
            \ee
            is a fundamental set of solutions of the equation $L_2\,y=0$.
            The application of the 
            \thref{fss} to the equation \eqref{eq1} gives us the following existence result.
            
            \bth{ffss}
             The equation \eqref{eq1} possesses a formal fundamental set of solution at the origin in the form
             $\{y_1(z), y_2(z), \psi_1(z), \psi_2(z)\}$ where the functions $y_j(z), j=1, 2$ are defines by \eqref{sol}.
             The functions $\psi_j(z), j=1, 2$ are given by
              \be\label{sol-0}\quad\quad
               \psi_1(z)=z^{-11/2}\,\exp \left(\frac{2 i\,\sqrt{3}}{3 z}\right)\,h_1(z),\quad
                  \psi_1(z)=z^{-11/2}\,\exp \left(-\frac{2 i\,\sqrt{3}}{3 z}\right)\,h_2(z)\,,
              \ee
              where
                  \ben
                  h_1(z)=z^6\left[-1+\frac{i\,\sqrt{3}}{6}\,z\right] +\frac{z^7}{12}\,\hat{\phi}(z),\quad
                  h_2(z)=z^6\left[-1-\frac{i\,\sqrt{3}}{6}\,z\right] +\frac{z^7}{12}\,\hat{\varphi}(z)\,.
                  \een
                  The elements $\hat{\phi}(z)$ and $\hat{\varphi}(z)$ are the following diveregent power series
                  \ben
                  \hat{\phi}(z)=\sum_{n=1}^{\infty}
                  \left(-\frac{i\,\sqrt{3}}{3}\right)^{n-1}\,\frac{(2\,n+1)!!}{2^{n-1}}\,z^n,\quad
                  \hat{\varphi}(z)=\sum_{n=1}^{\infty}
                  \left(\frac{i\,\sqrt{3}}{3}\right)^{n-1}\,\frac{(2\,n+1)!!}{2^{n-1}}\,z^n\,.
                  \een
              \ethe
              
              \proof
              We have only to prove that the functions $\psi_1(z)$ and $\psi_2(z)$ have the pointed form.
              From \thref{fss} it follows that the functions $\psi_j(z)$ must solve the equations
               \ben
                  & &
                \psi''(z) + \frac{2}{z}\,\psi'(z) + \frac{1}{3 z^4}\,\psi(z)=
                  z^{-11/2}\,\exp \left(\frac{2 i\,\sqrt{3}}{3 z}\right)\,\left[z^2 + \frac{i\,\sqrt{3}}{2}\,z^3\right],\\[0.2ex] 
                  & &
                   \psi''(z) + \frac{2}{z}\,\psi'(z) + \frac{1}{3 z^4}\,\psi(z)=
                   z^{-11/2}\,\exp \left(\frac{-2 i\,\sqrt{3}}{3 z}\right)\,\left[z^2 -\frac{i\,\sqrt{3}}{2}\,z^3\right]\,,  
               \een
               respectively.
                   Looking for a solution $\psi_1(z)$ of the first equation in the form
                  $$\,
                  \psi_1(z)=z^{-11/2}\,\exp\left(\frac{2\,i\,\sqrt{3}}{3 z}\right)\,h_1(z)\,,
                  \,$$
                 we find that $h_1(z)$ must solve the equation
                  $$\,
                  z^4\,h''_1(z) - \left[9 z^3 + \frac{4\,i\,\sqrt{3}}{3}\,z^2\right]\,h'_1(z)
                  +\left[\frac{99}{4}\,z^2 + \frac{22\,i\,\sqrt{3}}{3}\,z -1\right]\,h_1(z)=
                  z^6 + \frac{i\,\sqrt{3}}{2}\,z^7\,.
                  \,$$
                  Now it is not difficult to show that $h_1(z)$ has the pointed form. Similarly,
                  for the solution
                  $$\,
                  \psi_2(z)=z^{-11/2}\,\exp\left(-\frac{2\,i\,\sqrt{3}}{3 z}\right)\,h_2(z)
                  \,$$
                  of the second equation
                  we find that the  $h_2(z)$ must solve the equation
                  $$\,
                  z^4\,h''_2(z) - \left[9 z^3 - \frac{4\,i\,\sqrt{3}}{3}\,z^2\right]\,h'_2(z)
                  +\left[\frac{99}{4}\,z^2 - \frac{22\,i\,\sqrt{3}}{3}\,z -1\right]\,h_2(z)=
                  z^6 - \frac{i\,\sqrt{3}}{2}\,z^7\,.
                  \,$$
                  As before one can show that $h_2(z)$ has the poined form.
                  
                  This ends the proof.
              \qed
            
          Combining \thref{fss}, \thref{formal-0} and \thref{ffss}  we obtain the 
          formal fundamental matrix solution at the origin of the equation \eqref{eq1}.
          
          \bth{formal}
           The equation \eqref{eq1} possesses an unique formal fundamental matrix solution $\hat{\Phi}_0(z)$ at the
           origin in the form \eqref{fo}, where
            \be\label{lq}\qquad
             \Lambda=\diag \left(-2, -2, -\frac{15}{2}, - \frac{15}{2}\right),\quad
             Q=\diag\left(\frac{i\,\sqrt{3}}{3}, - \frac{i\,\sqrt{3}}{3}, \frac{2\,i\,\sqrt{3}}{3}, - \frac{2\,i\,\sqrt{3}}{3}\right)\,.
            \ee
            The matrix $\hat{H}(z)$ is given by
             \ben
              \hat{H}(z) &=& 
              \left(\begin{array}{cccc}
              	 z^2   & z^2  & z^2\,h_1(z)   & z^2\,h_2(z)\\[0.2ex]
              	 -\frac{i\,\sqrt{3}}{3}   & \frac{i\,\sqrt{3}}{3}   & [-\frac{11}{2}\,z - \frac{2\,i\,\sqrt{3}}{3}]\, h_1(z) + z^2\,h'_1(z)
              	 	& [-\frac{11}{2}\,z + \frac{2\,i\,\sqrt{3}}{3}]\, h_2(z) + z^2\,h'_2(z)\\[0.2ex]
              	0   & 0  & z^4 + \frac{i\,\sqrt{3}}{3} z^5  &  z^4 - \frac{i\,\sqrt{3}}{3} z^5\\[0.2ex]
              	0   & 0  & -\frac{2\,i\,\sqrt{3}}{3} z^2 -\frac{5 z^3}{2} - \frac{5 i\,\sqrt{3}}{4} z^4  
              	&  \frac{2\,i\,\sqrt{3}}{3} z^2 -\frac{5 z^3}{2} + \frac{5 i\,\sqrt{3}}{4} z^4  
              	               \end{array}\right)\\[0.15ex]
              	                 &=&
              	      \left(\begin{array}{cc}
              	      	\hat{H}_1(z)   & \hat{H}_{12}(z)\\
              	      	  0            & \hat{H}_2(z)
              	      	 \end{array}\right)           \,, 		
             \een
             where the functions $h_j(z), j=1, 2$ are defined by \thref{ffss}.
              \ethe
          
          \proof
             The proof is straightforward.
            \qed

  Once building a formal fundamental matrix solution at the origin we can compute the formal invariants of the
  equation \eqref{eq1}.
  
    \bpr{formal-inv}
     With respect to the formal fundamental matrix solution $\hat{\Phi}_0(z)$ at the origin built by \thref{formal}
     the exponential torus $\T$ and formal monodromy $\hat{M}_0$ of the equation \eqref{eq1} are given by
      $$\,
       \T=\left(\begin{array}{cccc}
         c  & 0        & 0    & 0\\
         0  & c^{-1}   & 0    & 0\\
         0  & 0        & c^2  & 0\\
         0  & 0        & 0    & c^{-2}
         \end{array}\right),\quad
         \hat{M}_0=\left(\begin{array}{ccrr}
         1  & 0        & 0    & 0\\
         0  & 1        & 0    & 0\\
         0  & 0        & -1   & 0\\
         0  & 0        & 0    & -1
         \end{array}\right)\,,
        \,$$
        where $c\in\CC^*$.
    \epr

      Now we have to compute the analytic invariants of the equation  \eqref{eq1}. 
      From \deref{sd} it follows that the only admissible singular direction $\theta$ associated to the divergent power series
      $\hat{\phi}(z)$ is  $\theta=\frac{\pi}{2}$. Similarly, the only admissible singular direction associated to the
      divergent power series $\hat{\varphi}(z)$ is $\theta=\frac{3 \pi}{2}$.
     The next lemma fixes explicitly the dependence of the pointed power series on the singular directions by providing the 
     1-summs of both  divergent power series $\hat{\phi}(z)$ and $\hat{\varphi}(z)$.

      \ble{sum}
         For any direction $\theta \neq \frac{\pi}{2}$ the function
          $$\,
           \phi_{\theta}(z)=3\,(-i\,\sqrt{3})^{-5/2}\int_0^{+\infty\,e^{i \theta}}
            \frac{e^{-\frac{\xi}{z}}}{\left(\xi -i\,\sqrt{3}\right)^{5/2}}\,d \xi
          \,$$      
          defines the 1-sum of the series $\hat{\phi}(z)$ in such a direction. Similarly, for any direction
          $\theta \neq \frac{3 \pi}{2}$ the function
           $$\,
           \varphi_{\theta}(z)=3\,(i\,\sqrt{3})^{-5/2}\int_0^{+\infty\,e^{i \theta}}
           \frac{e^{-\frac{\xi}{z}}}{\left(\xi + i\,\sqrt{3}\right)^{5/2}}\,d \xi
           \,$$      
           defines the 1-sum of the series $\hat{\varphi}(z)$ in such a direction. 
           
           The function $\phi_{\theta}(z)$ $($resp. $\varphi_{\theta}(z))$ is a holomorphic function
           in the open disk 
           \be\label{D}
           \D=\left\{z\in\CC\,|\,Re \left(\frac{e^{i \theta}}{z}\right) > 0 \right\}
           \ee
           for every $\theta\neq \frac{\pi}{2}$ $($resp. $\theta\neq \frac{3 \pi}{2})$.
         \ele            
  
     \proof
     
     Since $n+1 \leq 2^n$ when $n \in\NN$ then 
     \ben
     \left|\left(-\frac{i\,\sqrt{3}}{3}\right)^{n-1}\,\frac{(2 n+1)!!}{2^{n-1}}\right| 
         &<& 
     \left(\frac{\sqrt{3}}{3}\right)^{n-1} \,\frac{(2 n+2)!!}{2^{n-1}}=
     4\,\sqrt{3}\,\left(\frac{\sqrt{3}}{3}\right)^n\,(n+1)! \\[0.2ex]
        &\leq&
     4\,\sqrt{3}\,\left(\frac{2\,\sqrt{3}}{3}\right)^n\,n!\,.
     \een
     Hence both of the power series $\hat{\phi}(z)$ and $\hat{\varphi}(z)$ are Gevrey-1 series with constants
     $C=4\,\sqrt{3},\,A=\frac{2\,\sqrt{3}}{3}$.
   The corresponding formal Borel transforms 
       \ben
     \hat{\B}_1\,\hat{\phi} (\xi)
       &=&
       \sum_{n=0}^{\infty} \left(-\frac{i\,\sqrt{3}}{3}\right)^n
       \,\frac{(2 n+3)!!}{2^n}\,\frac{\xi^n}{n!},\\[0.25ex]
       \hat{\B}_1\,\hat{\varphi} (\xi)
       &=&
       \sum_{n=0}^{\infty} \left(\frac{i\,\sqrt{3}}{3}\right)^n
       \,\frac{(2 n+3)!!}{2^n}\,\frac{\xi^n}{n!}
           \een
    converge in the open disk $|\xi| < \sqrt{3}$ and there
     $$\,
     \hat{\B}_1\,\hat{\phi} (\xi)==3\left(1 + \frac{i\,\sqrt{3}}{3}\,\xi\right)^{-5/2}=\phi(\xi),\quad
     \hat{\B}_1\,\hat{\varphi}(\xi)=3\left(1 - \frac{i\,\sqrt{3}}{3}\xi\right)^{-5/2}=\varphi(\xi)\,.
     \,$$
     The function $\phi(\xi)$ (resp. $\varphi(\xi)$) is continued analytically along any  ray
     $\theta\neq \frac{\pi}{2}$ (resp. $\theta \neq \frac{3 \pi}{2}$) from $0$ to $+\infty e^{i \theta}$.
     Next, since
      \be\label{s1}
       |\phi(\xi)| \leq 
       \left\{\begin{array}{ccc}
       	 1   & \textrm{for}   & \sin \theta \leq 0,\\[0.15ex]
       	 1/|\cos \theta|^{5/2}   & \textrm{for}   & \sin \theta > 0
       	 \end{array} \right.
      \ee
      and
          \be\label{s2}
          |\varphi(\xi)| \leq 
          \left\{\begin{array}{ccc}
          	1   & \textrm{for}   & \sin \theta \geq 0,\\[0.15ex]
          	1/|\cos \theta|^{5/2}   & \textrm{for}   & \sin \theta < 0
          \end{array} \right.
          \ee
     the Laplace transform $(\L_1 \phi) (z)$ (resp. $(\L_1 \varphi (z)$) is well defined along any ray 
     $\theta \neq \frac{\pi}{2}$ (resp. $\theta \neq \frac{3 \pi}{2}$) from $0$ to $+ \infty e^{i \theta}$.
     Moreover, the estimates \eqref{s1} and \eqref{s2} ensure that the Laplace transforms
      \ben
       (\L_1 \phi) (z) &=&\phi_{\theta}(z)=
       3 \int_0^{+\infty e^{i \theta}}
       \frac{e^{-\frac{\xi}{z}}}{\left(1+ \frac{i\,\sqrt{3}}{3}\,\xi\right)^{5/2}}\,d \xi\\[0.3ex]
                &=&
        3\,(-i\,\sqrt{3})^{-5/2}\int_0^{+\infty\,e^{i \theta}}
         \frac{e^{-\frac{\xi}{z}}}{\left(\xi -i\,\sqrt{3}\right)^{5/2}}\,d \xi,\\[0.3ex]
          (\L_1 \varphi) (z)  &=& \varphi_{\theta}(z)=
         3 \int_0^{+\infty e^{i \theta}}
          \frac{e^{-\frac{\xi}{z}}}{\left(1 - \frac{i\,\sqrt{3}}{3}\,\xi\right)^{5/2}}\,d \xi\\[0.3ex]
            &=&
           3\,(i\,\sqrt{3})^{-5/2}\int_0^{+\infty\,e^{i \theta}}
             \frac{e^{-\frac{\xi}{z}}}{\left(\xi + i\,\sqrt{3}\right)^{5/2}}\,d \xi        
      \een
      define holomorphic functions in the open disk $\D$ from \eqref{D}.
      The so built Laplace  transforms define the 1-sums of the series $\hat{\phi}(z)$ and $\hat{\varphi}(z)$.
      
      This ends the proof.
     \qed

      \begin{figure}[t]
      	
      	\begin{minipage}[c][1\width]
      		{0.003\textwidth}
      		\centering
      		\includegraphics[scale=0.35,angle=180]{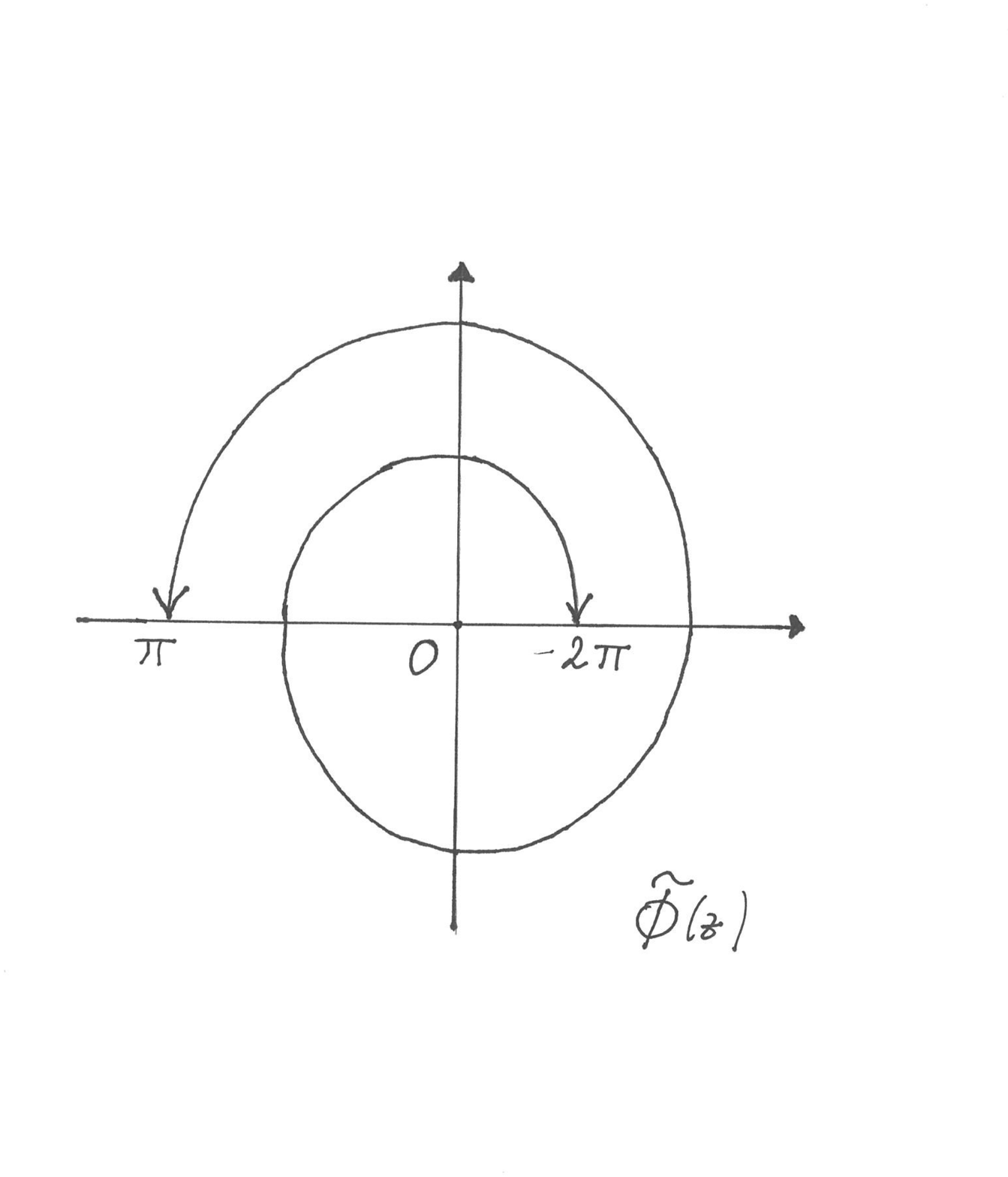}
      	\end{minipage}
      	\hfill
      	\begin{minipage}[c][1\width]
      		{0.5\textwidth}
      		\centering
      		\includegraphics[scale=0.35,angle=180]{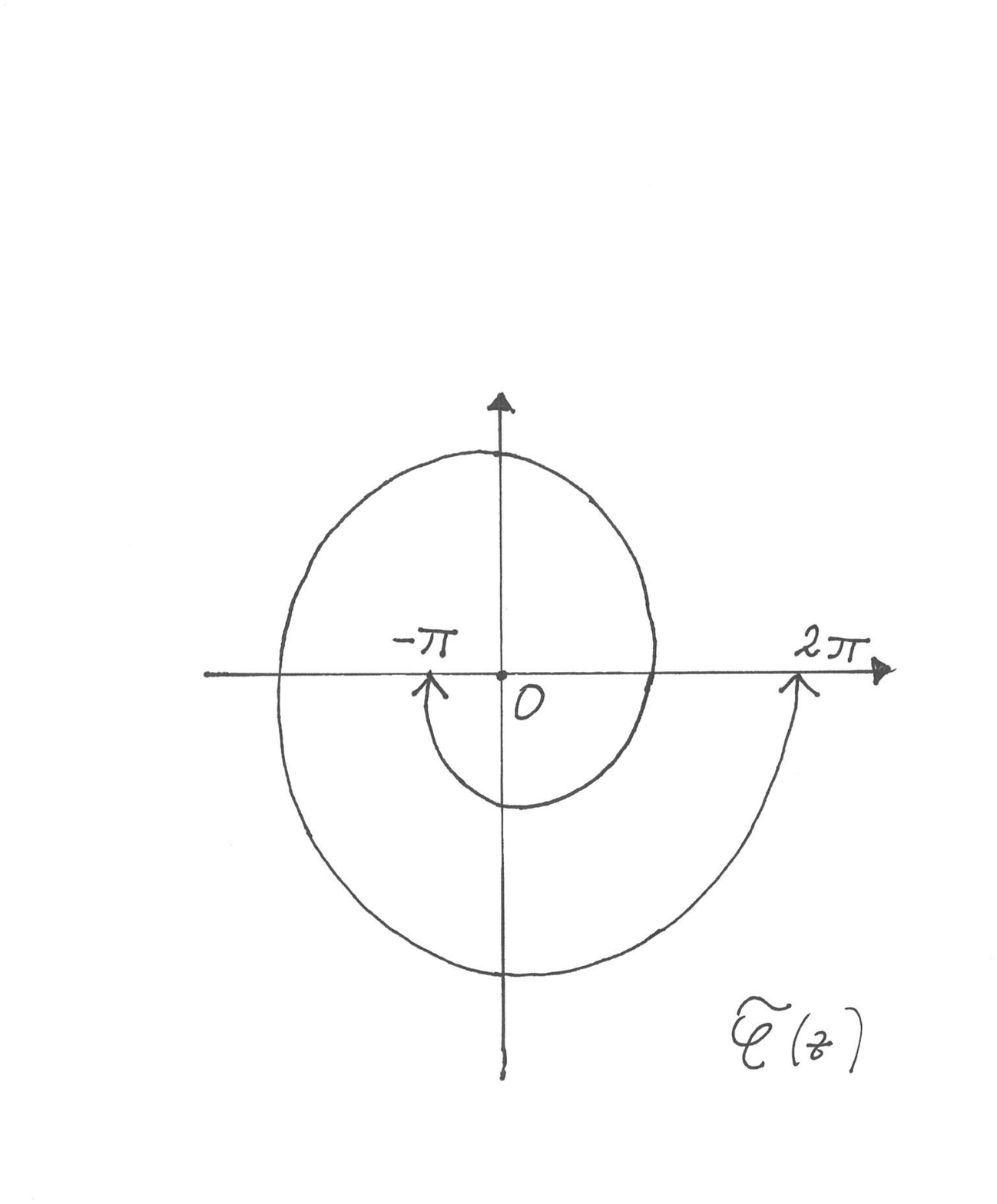}
      	\end{minipage}
      	\caption{Sectors $\tilde{D}$ of the Borel sums $\tilde{\phi}(z)$ and $\tilde{\varphi}(z)$.}
      	\label{fig:image}
      \end{figure}

     \bre{c}
      Let $I=\left(-\frac{3 \pi}{2}, \frac{\pi}{2}\right) \subset \RR$ and $J=\left(-\frac{\pi}{2}, \frac{3 \pi}{2}\right) \subset \RR$.
      When we move the direction $\theta\in I$ (resp. $\theta\in J$) the holomorphic functions $\phi_{\theta}(z)$
      (resp. $\varphi_{\theta}(z)$) glue together analytically and define a holomorphic function $\tilde{\phi}(z)$
      (resp. $\tilde{\varphi}(z)$) on a sector $\tilde{\D}$ with opening $3 \pi, \,-2 \pi < \arg (z) < \pi$
      (resp. $-\pi < \arg (z) < 2 \pi$) as it is shown in Figure \ref{fig:image}. On these sectors the functions $\tilde{\phi}(z)$ and $\tilde{\varphi}(z)$
      are asymptotic to the power series $\hat{\phi}(z)$ and $\hat{\varphi}(z)$, respectively, in Gevrey 1-sense and
      define the 1-sums of this power series there. The restriction of $\tilde{\phi}(z)$ (resp. $\tilde{\varphi}(z)$)
      on $\CC^*$ is a multivalued function. In every direction $\theta\neq \frac{\pi}{2}$ 
      (resp. $\theta \neq \frac{3 \pi}{2}$) the function $\tilde{\phi}(z)$ (resp. $\tilde{\varphi}(z)$)has only one
      values which coincides with the function $\phi_{\theta}(z)$ (resp. $\varphi_{\theta}(z)$) from \leref{sum}.
      Near the singular direction $\theta=\frac{\pi}{2}$ (resp. $\theta=\frac{3 \pi}{2}$) the function $\tilde{\phi}(z)$
      (resp. $\tilde{\varphi}(z)$) has two different values :
      $\phi^{+}_{\frac{\pi}{2}}(z)=\phi_{\frac{\pi}{2} + \epsilon}(z)$
      (resp. $\varphi^{+}_{\frac{3 \pi}{2}}(z)=\varphi_{\frac{3 \pi}{2}+\epsilon}(z)$) and
      	$\phi^{-}_{\frac{\pi}{2}}(z)=\phi_{\frac{\pi}{2} - \epsilon}(z)$
      	(resp. $\varphi^{-}_{\frac{3 \pi}{2}}(z)=\varphi_{\frac{3 \pi}{2}-\epsilon}(z))$ for a small number $\epsilon > 0$.
     \ere
     
     Now we already can associate to the formal fundamental matrix solution $\hat{\Phi}_0(z)$ from
     \thref{formal} an actual fundamental matrix solution in the sense of \thref{R}.
     Denote $F(z)=z^{\Lambda}\,\exp\left(\frac{Q}{z}\right)$. 
     
     \bth{actual}
      For every direction $\theta\neq \{\frac{\pi}{2}, \frac{3 \pi}{2}\}$ the equation \eqref{eq1} possesses an unique
      actual fundamental matrix solution $\Phi_0^{\theta}(z)$ at the origin in the form
       \be\label{F}
        \Phi^{\theta}_0(z)=H_{\theta}(z)\,F_{\theta}(z)\,,
       \ee
       where $F_{\theta}(z)$ is the branch of the matrix $F(z)$ for $\arg (z)=\theta$. The matrix $H_{\theta}(z)$ is given by
        \ben
          H_{\theta}(z)=\left(\begin{array}{cc}
          	H_1(z)   & H_{12}^{\theta}(z)\\
          	  0               & H_2(z)
          	\end{array}\right)\,,
        \een
        where $H_j(z)=\hat{H}_j(z), j=1, 2$. The matrix $H_{12}^{\theta}(z)$ is defined as
         \ben
          H_{12}^{\theta}(z)=\left(\begin{array}{cc}
          	  z^2\,h_1^{\theta}(z)    & z^2\,h_2^{\theta}(z)\\
          	  (\frac{11}{2} z - \frac{2 i\,\sqrt{3}}{3}) h_1^{\theta}(z) + z^2 \frac{d h_1^{\theta}(z)}{d z}
          	   &  	  (\frac{11}{2} z + \frac{2 i\,\sqrt{3}}{3}) h_2^{\theta}(z) + z^2 \frac{d h_2^{\theta}(z)}{d z}
          	\end{array}\right)\,.
         \een
         The functions $h_j^{\theta}(z), j=1, 2$ are given by
         $$\,
            h_1^{\theta}(z)=z^6\left[-1+\frac{i\,\sqrt{3}}{6}\,z\right] +\frac{z^7}{12}\,\phi_{\theta}(z),\quad
            h_2^{\theta}(z)=z^6\left[-1-\frac{i\,\sqrt{3}}{6}\,z\right] +\frac{z^7}{12}\,\varphi_{\theta}(z)\,,
         \,$$
         where the functions $\phi_{\theta}(z)$ and $\varphi_{\theta}(z)$ are defined by \leref{sum} and extended by \reref{c}.

         Near the singular direction $\theta=\frac{\pi}{2}$ the equation \eqref{eq1} possesses two different fundamental matrix solutions 
         at the origin          
           $$\,
           \Phi^{\pi/2+\epsilon}_0(z)
            \quad \textrm{and}\quad \Phi^{\pi/2-\epsilon}_0(z)\,,
                        \,$$
         where $\Phi_0^{\pi/2+\epsilon}(z)$ and $\Phi_0^{\pi/2-\epsilon}(z)$ are defined by \eqref{F} for a small number
         $\epsilon > 0$.               

         Similarly, near the singular direction $\theta=\frac{3 \pi}{2}$ the equation \eqref{eq1}
          possesses two different fundamental matrix solutions 
          at the origin          
          $$\,
          \Phi^{3 \pi/2+\epsilon}_0(z)
          \quad \textrm{and}\quad
          \Phi^{3 \pi/2-\epsilon}_0(z)\,,
          \,$$
          where $\Phi_0^{3 \pi/2+\epsilon}(z)$ and $\Phi_0^{3 \pi/2-\epsilon}(z)$ are defined by \eqref{F} for a small number
          $\epsilon > 0$.               
     \ethe

  Now we are ready to compute explicitly the Stokes matrices at the origin of the equation \eqref{eq1}.

   \bth{stokes}
    With respect to the actual fundamental matrix solution at the origin defined by \thref{actual} the equation
    \eqref{eq1} has a Stokes matrix $St_{\pi/2}$ in the form
     \ben
      St_{\pi/2}=\left(\begin{array}{cccc}
      	 1   & 0   & 0      & 0\\
      	 0   & 1   & \mu_1  & 0\\
      	 0   & 0   & 1      & 0\\
      	 0   & 0   & 0      & 1
      	\end{array}\right)\,,
     \een
     where
      $$\,
       \mu_1=\frac{2\,(-i\,\sqrt{3})^{5/2}\,\sqrt{\pi}}{3}\,.
      \,$$
      Similarly,  respect to the actual fundamental matrix solution at the origin defined by \thref{actual} the equation
      \eqref{eq1} has a Stokes matrix $St_{3 \pi/2}$ in the form
      \ben
      St_{3 \pi/2}=\left(\begin{array}{cccc}
      	1   & 0   & 0      & \mu_2\\
      	0   & 1   & 0      & 0\\
      	0   & 0   & 1      & 0\\
      	0   & 0   & 0      & 1
      \end{array}\right)\,,
      \een
      where
      $$\,
      \mu_2=\frac{2\,(i\,\sqrt{3})^{5/2}\,\sqrt{\pi}}{3}\,.
      \,$$
   \ethe

   \proof

  From \deref{stokes} it follows that  to find the  multiplier $\mu_1$ we have to compare the solutions
   $$\,
   \left[\psi_1(z)\right]^{-}_{\pi/2}=z^{1/2}\,\exp\left(\frac{2 i\,\sqrt{3}}{3 z}\right)\,
   	\left[-1 +\frac{i \,\sqrt{3}}{6}\,z\right]+
   	\frac{z^{3/2}}{12}\,\exp\left(\frac{2 i\,\sqrt{3}}{3 z}\right)\,\phi^{-}_{\pi/2}(z)
   	\,$$	
   		and
   	 $$\,
   	 \left[\psi_1(z)\right]^{+}_{\pi/2}=z^{1/2}\,\exp\left(\frac{2 i\,\sqrt{3}}{3 z}\right)\,
   	 	\left[-1 +\frac{i \,\sqrt{3}}{6}\,z\right]+
   	 		\frac{z^{3/2}}{12}\,\exp\left(\frac{2 i\,\sqrt{3}}{3 z}\right)\,\phi^{+}_{\pi/2}(z)\,.
   	 			\,$$			
   	Then 			
   
    \ben
     \left[\psi_1(z)\right]^{-}_{\pi/2} -  \left[\psi_1(z)\right]^{+}_{\pi/2}
      &=&
      \frac{1}{12}\,z^{3/2}\,\exp \left(\frac{2 i\,\sqrt{3}}{3 z}\right)
      \left[\phi^{-}_{\pi/2}(z) - \phi^{+}_{\pi/2}(z)\right]\\[0.25ex]
       &=&
      \frac{(-i\,\sqrt{3})^{5/2}}{4}\,z^{3/2}\,\exp \left(\frac{2 i\,\sqrt{3}}{3 z}\right)\,
      \int_{\gamma} (\xi-i\,\sqrt{3})^{-5/2}\,\mathrm{e}^{-\frac{\xi}{z}}\,
      \mathrm{d}\xi \,,
    \een
    where $\gamma=(\pi/2-\epsilon) - (\pi/2 + \epsilon)$. Without changing the integral we can deform the path $\gamma$
    into  a Henkel type path going along $i\,\RR^{+}$ from $i\,\infty$ to $i\,\sqrt{3}$, encircling $i\,\sqrt{3}$
    in the positive sense and backing to $i\,\infty$. Then
     \ben
       & &
        \frac{(-i\,\sqrt{3})^{5/2}}{4}\,z^{3/2}\,\exp \left(\frac{2 i\,\sqrt{3}}{3 z}\right)\,
      \left(1-\mathrm{e}^{-2 \pi\,i (-5/2)}\right)
      \int_{i\,\sqrt{3}}^{i\,\infty}
      (\xi -i\,\sqrt{3})^{-5/2} \,\mathrm{d}\xi\\[0.25ex]
        &=&
          \frac{(-i\,\sqrt{3})^{5/2}}{2}\,z^{3/2}\,\exp \left(-\frac{i\,\sqrt{3}}{3 z}\right)\,
          \int_0^{i\,\infty}
          u^{-5/2}\,\mathrm{e}^{-5/2}\,\mathrm{d} u\\[0.25ex]
          &=&
            \frac{(-i\,\sqrt{3})^{5/2}}{2}\,\exp \left(-\frac{i\,\sqrt{3}}{3 z}\right)\,
            \int_0^{+\infty}
            \tau^{-5/2}\,\mathrm{e}^{-\tau}\,\mathrm{d} \tau\\[0.25ex]
          &=&
            \frac{(-i\,\sqrt{3})^{5/2}}{2}\,\exp \left(-\frac{i\,\sqrt{3}}{3 z}\right)\,
            \Gamma\left(-\frac{3}{2}\right)=
             \frac{(-i\,\sqrt{3})^{5/2}}{2}\,
             \frac{\pi}{\Gamma\left(1+ \frac{3}{2}\right)}
             \exp \left(-\frac{i\,\sqrt{3}}{3 z}\right)\\[0.25ex]
               &=&
                \frac{2\,(-i\,\sqrt{3})^{5/2}\,\sqrt{\pi}}{3}\,
                \exp \left(-\frac{i\,\sqrt{3}}{3 z}\right)=
                \mu_1\,y_2(z)\,,  
            \een
       where we have used the Euler's reflection formula $\Gamma(1-z)\,\Gamma(z)=\frac{\pi}{\sin \pi\,z}$ for $z\neq \ZZ$.
       In the same manner we find that
        \ben
        \left[\psi_2(z)\right]^{-}_{3 \pi/2} -  \left[\psi_2(z)\right]^{+}_{3 \pi/2}
        &=&
            \frac{1}{12}\,z^{3/2}\,\exp \left(-\frac{2 i\,\sqrt{3}}{3 z}\right)
            \left[\varphi^{-}_{3\pi/2}(z) - \varphi^{+}_{3 \pi/2}(z)\right]\\[0.25ex]
            &=&
            \frac{2\,(i\,\sqrt{3})^{5/2}\,\sqrt{\pi}}{3}\,
            \exp \left(\frac{i\,\sqrt{3}}{3 z}\right)=
            \mu_2\,y_1(z)\,.     
        \een

   This ends the proof. \qed
   
   Thanks to \prref{formal-inv} and \thref{stokes} we can describe the local differential Galois group at the origin
   of the equation \eqref{eq1}.
   
   \bth{G}
    The connected component of the unit element of the local differential Galois group at the origin of the equation
    \eqref{eq1} is not Abelian.
   \ethe 
   
   \proof

   From \prref{formal-inv} it follows that the group generated by $\T$ and $\hat{M}_0$ is not a connected group.
   However the connected component of the unit element of this group is equal to the exponential torus $\T$.
   From the \thref{Ramis} of Ramis it follows that the local differential Galois group at the origin of the equation
   \eqref{eq1} is the Zariski closure of the subgroup generated by the exponential torus $\T$ and the Stokes matrices
   $St_{\pi/2}$ and $St_{3 \pi/2}$. Denote by $S$ the Zariski closure of the subgroup generated by the
   Stokes matrices $St_{\pi/2}$ and $St_{3 \pi/2}$. Then the element $s_{\lambda, \nu}$ of $S$ has the form
    $$\,
      s_{\lambda, \nu}=\left(\begin{array}{cccc}
       1   & 0   & 0    & \lambda\\
       0   & 1   & \nu  & 0\\
       0   & 0   & 1    & 0\\
       0   & 0   & 0    & 1
       \end{array}\right)\,,
    \,$$   
    where $\lambda, \nu\in\CC$. Denote by $T$ the Zariski closure of the subgroup generated by the exponential
    torus $\T$. Then the element $\tau_c$ of $T$ has the form
     $$\,
       \tau_c=\left(\begin{array}{cccc}
         c  & 0       & 0    & 0\\
         0  & c^{-1}  & 0    & 0\\
         0  & 0       & c^2  & 0\\
         0  & 0       & 0    & c^{-2}
          \end{array}\right)\,,
     \,$$
     where $c\in\CC^*$.
   
    When $\lambda \neq 0, \nu \neq 0$ and  $c^3 \neq 1$  the commutator between $s_{\lambda, \nu}$
    and $\tau_c$
      $$\,
       s_{\lambda, \nu}\,\tau_c\,s^{-1}_{\lambda, \nu}\,\tau^{-1}_c=
       \left(\begin{array}{cccc}
         1   & 0  & 0                  & (1-c^3)\,\lambda\\
         0   & 1  & (1-c^{-3})\,\nu    & 0\\
         0   & 0  & 1                  & 0\\
         0   & 0  & 0                  & 1
         \end{array}\right)
      \,$$
      is not identically equal to the identity matrix.
      
      The condition $c^3=1$ implies that the group generated by $\T$ is a finite cyclic group. For such a group the
      Picard-Vessiot extension $K\left(e^{\frac{i \,\sqrt{3}}{3 z}}\right)$ must be an algebraic extension of
      the field $K=\CC(z)$  which is an obvious contradiction. Thus the connected
      component of the unit element of the local differential Galos group at the origin of the equation \eqref{eq1} is
      not Abelian.
      
      This ends the proof.
   \qed

   \vspace{3ex}
   
   Now we will determine the local differential Galois group at $z=\infty$ of the equation \eqref{eq1}.
   The transformation $z=1/x$ takes the equation \eqref{eq1}  into the equation
    \be\label{eq2}
     \frac{d^4 y}{d x^4} + \frac{1}{x}\,\frac{d^3 y}{d x^3} 
     +\left[-\frac{9}{4 x^2} + \frac{5}{3}\right]\,\frac{d^2 y}{d x^2} +
     \frac{1}{3 x}\,\frac{d y}{d x} - \left[\frac{3}{4 x^2} - \frac{4}{9}\right]\,y=0\,.
    \ee    
      The origin is a regular singularity for the equation \eqref{eq2} and therefore the point
      $z=\infty$ is a regular singularity for the equation \eqref{eq1}. 
      The characteristic equation at $x=0$ for the equation \eqref{eq2} writes
       $$\,
        \rho\,(\rho-1)\,\left(\rho^2 -4 \rho + \frac{7}{4}\right)=0\,.
       \,$$
      Its roots are $\rho_1=0, \rho_2=1, \rho_3=\frac{1}{2}, \rho_4=\frac{7}{2}$ and therefore the solution
      at the origin can contain logarithmic terms. Fortunately the reducibility of the equation \eqref{eq2} allows us to
      overcome the technical difficulty in the building of a local fundamental set of solutions at such a 
      resonant regular singularity. In fact, as we show below the solution at the origin of the equation
      \eqref{eq2} does not contain logarithmic terms. 
      
      The operators $L_1$ and $L_2$ for the equation \eqref{eq2}
      become
       $$\,
        L_1=\partial^2 + \frac{1}{3}, \qquad
        L_2=\partial^2 + \frac{1}{x}\,\partial - \frac{9}{4 x^2} + \frac{4}{3}\,.
       \,$$
       The origin is an ordinary point for the equation $L_1\,y=0$. In fact, the function
       $y_1(x)=\exp\left(\frac{i\,\sqrt{3}\,x}{3}\right)$ and
       $y_2(x)=\exp\left(-\frac{i\,\sqrt{3}\,x}{3}\right)$ form a fundamental set of solutions of the
       equation $L_1\,y=0$. The following lemma gives a local fundamental set of solutions near the origin
       for the equation $L_2\,y=0$.
       
       \ble{2-s}
        The equation $L_2\,y=0$ possesses a local fundamental set of solutions near the origin in the form
         $$\,
          w_1(x)=x^{3/2}\,u_1(x), \qquad w_2(x)=x^{-3/2}\,u_2(x)\,,
         \,$$
         where $u_j(x), j=1, 2$ are holomorphic functions near the origin. In particular,
         $$\,
          u_1(x)=\sum_{k=0}^{\infty} a_{2 k}\,x^{2 k},\quad a_0=1,\qquad
          u_2(x)=\sum_{k=0}^{\infty} b_{2 k}\,x^{2 k}, \quad b_0=1\,,
         \,$$
         where the coefficients $a_{2 k}$ and $b_{2 k}$ satisfy the recurrence relations
         \be\label{rec}
          3 k\,(2 k+3)\,a_{2 k} + 2 a_{2 k -2}=0,\qquad
          3 k \,(2 k-3)\,b_{2 k} + 2 b_{2 k-2}=0\,,
         \ee
         respectively.
       \ele
       
       \proof
       The characteristic equation at $x=0$ for the equation
        $$\,
         L_2\,y=y'' + \frac{1}{x}\,y' + \left[-\frac{9}{4 x^2} + \frac{4}{3}\right]\,y=0
        \,$$
        writes
         $$\,
          \rho\,(\rho-1) + \rho - \frac{9}{4}=0\,.
         \,$$
         Its roots are $\rho_1=-\frac{3}{2}$ and $\rho_2=\frac{3}{2}$. The local theory of regular (Fuchsian)
         singularity ensures that the equation $L_2\,y=0$ admits a local solution $w_1(x)$ near the origin in the
         form $w_1(x)=x^{3/2}\,u_1(x)$ where $u_1(x)$ is a holomorphic function near the origin. The function
         $u_1(x)$ must satisfies the equation
          $$\,
           3 x\,u''_1(x) + 12\,u'_1(x) + 4 x\,u_1(x)=0\,.
           \,$$
           Obviously the wanted function $u_1(x)$ is an even function in $x$. Looking $u_1(x)$ as a power series
           $$\,
            u_1(x)=\sum_{k=0}^{\infty} a_{2 k}\,x^{2 k}=a_0 + a_2\,x^2 + a_4\,x^{2 k} + \cdots + a_{2 n-2}\,x^{2 n-2} +
            a_{2 n}\,x^{2 n} + \cdots
            \,$$ 
        we find that the coefficients $a_{2 k}$ must satisfy the recurrence relation \eqref{rec}. 
        
        Since $\rho_2-\rho_1\in\NN$ the second solution $w_2(x)$ at the origin can contain logarithmic term. It turns out that
        the evenness of the function $u_1(x)$ forbids the existence of the logarithmic term. Indeed, 
        the solution $w_2(x)$ depends on the solution $w_1(x)$ by the formula (see \cite{G} for details)
         $$\,
          w_2(x)=w_1(x)\,\int \frac{e^{-\int \frac{d x}{x}}}{w^2_1(x)}\,d x=
          w_1(x)\int \frac{d x}{x^4\,u^2_1(x)}\,.
         \,$$
         Since the function $u_1(x)$ is an even function, so the function $1/u^2_1(x)$. Therefore the solution
         $w_2(x)$ does not contain logarithmic term. Thus the solution $w_2(x)$ has the form
         $w_2(x)=x^{-3/2}\,u_2(x)$ where $u_2(x)$ must solve the equation
          $$\,
           3 x\,u''_2(x) - 6\,u'_2(x) + 4 x\,u_2(x)=0\,.
          \,$$
          As above we see that the function $u_2(x)$ is an even function 
           $$\,
           u_2(x)=\sum_{k=0}^{\infty} b_{2 k}\,x^{2 k}=b_0 + b_2\,x^2 + b_4\,x^{2 k} + \cdots + b_{2 n-2}\,x^{2 n-2} +
           b_{2 n}\,x^{2 n} + \cdots
           \,$$ 
           and the coefficients $b_{2 k}$ must satisfy the recurrence relation \eqref{rec}.
           
           This ends the proof.   
       \qed

      \bth{fss-in}
        The equation \eqref{eq2} possesses a fundamental set of solutions near the origin in the form
        $\{y_1(x), y_2(x), \psi_1(x), \psi_2(x)\}$ where $y_j(x), j=1, 2$ are holomorphic functions near the
        origin. The solutions $\psi_1(x)$ and $\psi_2(x)$ write
         $$\,
          \psi_1(x)=x^{7/2}\,v_1(x),\qquad 
          \psi_2(x)=x^{1/2}\,v_2(x)\,,
         \,$$
         where $v_j(x), j=1, 2$ are holomorphic functions near the origin. In particular
         $$\,
          v_1(x)=\sum_{k=0}^{\infty} c_{2 k}\,x^{2 k},\quad c_0=\frac{4}{35},\qquad
          v_2(x)=\sum_{k=0}^{\infty} d_{2 k}\,x^{2 k}, \quad d_0=-\frac{1}{4}\,.
         \,$$
         The coefficients $c_{2 k}$ and $d_{2 k}$ satisfy  the recurrence relations
          \be\label{rec-1}
           & &
           (48 k^2 + 48 k +9)\,c_{2 k} + 4 c_{2 k -2}=12\,a_{2k-2},\\[0.15ex]
            & &
           (48 k^2 -96 k +45)\,d_{2 k} + 4 d_{2k-2}=12\,b_{2k-2}\,,\nonumber
          \ee
          where $a_{2 k}$ and $b_{2 k}$ are defined by \leref{2-s}.
      \ethe
      
      \proof
      
      From \thref{fss} it follows that the solutions $\psi_1(x)$ and $\psi_2(x)$ must satisfy the equations
       \ben
        \psi''_1(x) + \frac{1}{3}\,\psi_1(x)=x^{3/2}\,u_1(x)\,,\quad
        \psi''_2(x) + \frac{1}{3}\,\psi_2(x)=x^{-3/2}\,u_2(x)\,,
       \een
       respectively, where $u_j(x)$ are defined by \leref{2-s}. Looking $\psi_j(x), j=1, 2$
       as $\psi_1(x)=x^{3/2}\,v_1(x)$ and $\psi_2(x)=x^{-3/2}\,v_2(x)$ we find that $v_j(x), j=1, 2$ must satisfy the
       equations
        \ben
         & &
         12\,x^2\,v''_1(x) + 36\,x\,v'_1(x) + (9 + 4\,x^2)\,v_1(x) = 12 \,x^2\,u_1(x),,\\[0.15ex]
          & &
        12\,x^2\,v''_2(x) - 36\,x\,v'_2(x) + (45 + 4\,x^2)\,v_2(x) = 12 \,x^2\,u_2(x),, 
        \een
        respectively. As above we see that the functions $v_1(x)$ and $v_2(x)$ must be even functions.
        Looking for them as power series in $x^2$ we find the recurrence  relations \eqref{rec-1} for
        their coefficients.
        
        This ends the proof.\qed

       Denote by $\Phi_0(x)$ a fundamental matrix solution at the origin of the equation \eqref{eq2}.
       Then from \thref{fms} it follows that
       \bth{origin}
        The fundamental matrix solution $\Phi_0(x)$ of the equation \eqref{eq2} is represented in a neighborhood
        of the origin as
         $$\,
          \Phi_0(x)=Y(x)\,x^{J}\,,
         \,$$
         where the matrix $Y(x)$ is a holomorphic matrix function there.
         In particular $Y(x)$ is defined as
         \ben
         Y(x)=\left(\begin{array}{cccc}
         y_1(x)   & y_2(x)   & x^3\,v_1(x)   & x^3\,v_2(x)\\[0.1ex]
         y'_1(x)  & y'_2(x)	 & \frac{7}{2}\,x^2\,v_1(x) + x^3\,v'_1(x)  
         & \frac{1}{2}\,x^2\,v_2(x) + x^3\,v'_2(x)\\[0.15ex]
         0        & 0        & x\,u_1(x)    & x\,u_2(x)\\[0.1ex]
         0        & 0        & \frac{3}{2}\,u_1(x) + x\,u'_1(x)
         & -\frac{3}{2}\,u_2(x) + x\,u'_2(x)
                \end{array}\right)\,,
         \een
         where the functions $u_j(x)$ and $v_j(x), j=1,2$ are defined by \leref{2-s} and \thref{fss-in}, respectively.
         The matrix $J$ is given by
          $$\,
           J=\diag\left(0, 0, \frac{1}{2}, -\frac{5}{2}\right)\,.
          \,$$
       \ethe

       Now we can describe the local differential Galos group at the origin of the equation \eqref{eq2}.
       
        \bth{m}
         The connected component of the unit element of the local differential Galois group at the origin of the equation \eqref{eq2} is trivial.
        \ethe
         
    \proof
    
       With respect to the fundamental matrix solution $\Phi_0(x)$ fixed by \thref{origin} the equation
     \eqref{eq2} has a monodrmy matrix $M_0$ around the origin in the form
     $$\,
     M_0=\diag (1, 1, -1, -1)\,.
     \,$$
     From \thref{Sh} of Schlesinger it follows that the local differential Galois group of equation \eqref{eq2} is 
     generated topologically by the monodromy matrix $M_0$. This group is not connected.
        However, the connected
        component of the unit element of this group is trivial. 
        
        This ends the proof.
    \qed

        Thanks to \thref{G} and \thref{m} we establish the main result of this section
        
        \bth{b}
          Assume that $a=0$ and $b=-\frac{2}{9}$. Then the fourth Painlev\'{e} equation is not integrable in
          the Liouville-Arnold sense by meromorphic functions that are rational in $t$.      
        \ethe

        \proof
         
        From \thref{m} it follows that  the local differential Galois group at $z=\infty$ of the
        equation \eqref{eq1} can be considered as  a subgroup of the local differential Galois group at the origin
        of the same equation. Then form \thref{G} and \thref{Mitschi}  of Mitschi it follows that the connected component
        of the unit element of the global differential Galois group of the equation \eqref{eq1} is not Abelian.
        Thus the Hamiltonian system \eqref{h} is not integrable.
        
        This ends the proof.\qed

    \section{Proof of \thref{main}}
    
    With the exception of the first Painlev\'{e} equation all of the Painlev\'{e} equations admits groups of symmetries 
    called auto-B\"{a}cklund transformations. The auto-B\"{a}cklund transformations relate one solution of a given
    Painlev\'{e} equation to another solution of the same equation, possibly with different values of the parameters
    (see \cite{BCH, C}). Using the auto-B\"{a}cklund transformations one can obtain various property of the Painlev\`{e}
    equations, including hierarchies of exact solutions \cite{BCH}, special integrals (see \cite{Gr}). 
    In this section applying auto-B\"{a}cklund transformations we extend the non-integrable result obtained in
    the previous section.

      From the works of Bassom et. al \cite{BCH} and Murata \cite{M} it follows that 
      the hierarchy stirred from the rational solution (II) by auto-B\"{a}cklund transformations has the form
      $$\,
      y(t; a_n, b_n)=-\frac{2}{3}\,t + \frac{P_{n-1}(t)}{Q_n(t)}\,,
      \,$$
      where $P_{n-1}(t)$ and $Q_n(t)$ are polynomials of degree $n-1$ and $n$, respectively, for the parameters
      $$\,
      (a_n, b_n)=\left(m, -\frac{2}{9} (1+3 n)^2\right),\quad m, n\in\ZZ,\quad \frac{1}{2} (m+n)\in\ZZ\,.
      \,$$       
           
         {\it Proof of the \thref{main}}.
         The proof follows from the fact that the auto-B\"{a}cklund transformations are birational canonical transformations
         \cite{O, Wa, ZF}.
         \qed

      		\vspace{1cm}
      		{\bf Acknowledgments.}\,
     The author was partially supported by Grant KP-06-N 62/5  of the Bulgarian Found 
    "Scientific research".

\begin{small}
    
\end{small}

\end{document}